# Preprint Extending Touch-less Interaction on Vision Based Wearable Device

Zhihan Lv*
SIAT, Chinese Academy of Science,
518055, Shenzhen, PRC.

Shengzhong Feng
SIAT, Chinese Academy of Science,
518055, Shenzhen, PRC.

Liangbing Feng
SIAT, Chinese Academy of Science,
518055, Shenzhen, PRC.

Haibo Li
Royal Institute of Technology (KTH),
10044, Stockholm, Sweden.

**ABSTRACT**

A touch-less interaction technology on vision based wearable device is designed and evaluated. Users interact with the application with dynamic hands/feet gestures in front of the camera. Several proof-of-concept prototypes with eleven dynamic gestures are developed based on the touch-less interaction. At last, a comparing user study evaluation is proposed to demonstrate the usability of the touch-less approach, as well as the impact on user's emotion, running on a wearable framework or Google Glass.

**Index Terms:** H.1.2 [User/Machine Systems]: Human factors—; H.5.1 [Multimedia Information Systems]: Artificial, augmented, and virtual realities—

## 1 INTRODUCTION

Nowadays, there is an increasing interest in creating wearable device interaction approach. Novel emerging user interface technologies have the potential to significantly affect market share in PC, smartphones, tablets and latest wearable devices such as head-wearable device(HWD), i.e. Google Glass, since the miniaturization of mobile computing devices permits 'anywhere' access to the information [8]. Therefore, displacing these technologies in smart devices is becoming a hot topic. Google glass has many impressive characteristics, and will not meet the occlusion problem and the fat finger problem [12], which frequently occurs in direct touch controlling mode, anymore. However, Google Glass only provides a touch pad that includes haptic with simple *tapping and sliding your finger* gestures, which is a one-dimensional (1D) interaction in fact and limits the intuitive and flexibility of interaction. Therefore, implementing a light-weight touch-less gesture recognition vision systems could provide a easy-to-use interaction approach. It has been proven effectiveness in virtual environments [11].

This paper researches the potential of the touch-less approach on two vision based wearable devices: a hybrid wearable framework where users mount smart phone over wrist or knee, and HWD, i.e. Google Glass. Four touch-less applications were developed. The questionnaires are designed to evaluate the usability of the eleven dynamic hand/foot gestures, and emotions of the users. The aim of our work is to identify the good design guidelines and options for interactive interface of vision based wearable devices. To focus on the actual interaction rather than computer vision related problems, we therefore adopt a successful dynamic interaction algorithm based on our previous research on foot [1] and on hand [4], the demonstrations are derived from [3]. The three dimensional interaction and stereoscopic visualization has been explored but not ported on wearable device yet due to the high computing consumption [6].

---

*e-mail: lvzhihan@gmail.com

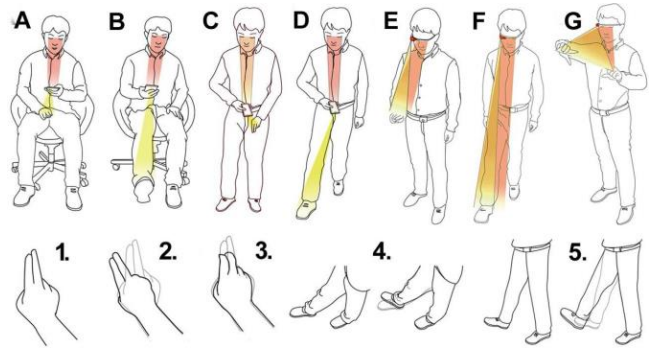

Figure 1: From up-to-down: The touch-less interaction on either smartphone or smart glasses; Hands and feet gestures.

## 2 SYSTEM

When the user wears the hardware, the software tracks the finger/foot dynamic gestures. Figure 1 down shows the basic gestures of hands $(1,2,3)$ and feet $(4,5)$. 1 shows the initial hands' gesture and orientation, 2 illustrates that the finger swings from left to right(or opposite direction). Besides defining the direction by swinging fingers, users can also move the visual mouse by moving their hands. 3 represents the flex and extension motion of fingers, which is similar to the gesture of clicking mouse. 4 indicates the motion of feet, which can trigger some events related to the position in certain scene by tiptoe. 5 illustrates the forward motion which can trigger optional events kick and moving motion according to the moving speed of users feet.

Figure 1 up-left shows that, regardless of the body gesture(sitting as in $A$, $B$, standing as in $C$, $D$), the device can recognize the hand or feet gesture via the rear camera, according to which the device can manipulate the software. Red zone indicates the users sight, yellow zone indicates the cameras video capture zone. Apparently, the distance from eye to screen has no difference, but the distance from the camera to hand $(A,C)$ is clearly shorter than to feet $(B,D)$. Moreover, the captured gesture via the rear camera is displayed through the screen, presenting to the users indirectly. While for those gestures via the front camera, users can see the visual react directly. Furthermore, one hand has to hold the hand-held device $(A,B,C,D)$, which will weaken the interaction efficiency. The Figure 1 up-right illustrates the user interacts with the glasses which can recognize the gestures. The users sight indicated by red zone basically coincides with the camera capturing zone indicated by yellow zone, because the eyes are close to the camera, and they have the same orientation. The gestures seen by user are generally the same with those captured by the camera, which makes this kind of interaction via wearable glasses is more intuitive than that with the mobile. Also, it frees the users hands totally so that operating the device with both hands is well-supported (shown as $G$).

## 3 EVALUATION

Four proof-of-concept wearable demonstrations to illustrate and explore different use cases, presented at [3]. Accordingly, we report on a user study comparing usability and emotions of the users. We expect the user to perceive the touch-less approach as being stretched from his/her hand/foot for manipulating content on the screen. In order to complete this interaction loop, our system maps and updates the physical hand/foot onto the screen coordinates. The system has been presented at Siggraph Asia 2014 [5].

Two user groups were participated into the experiment for the comparing research. The first user group consisting of 15 participants (6 female), from a diverse background aged from 25-46 (m=30.13, sd=5.74). They tested to interact with the apps on the hybrid wearable framework and answered some open-ended questions. A two-stages questionnaire is required to answer to elicit participants' reactions: Usability of the Gestures and User Emotions. Usability describes the quality of user experience of software and its interaction. Emotion is a significant part of user's decision-making ability. Emotion testing can evaluate the product against different emotions that a user goes through. The second user group includes 15 participants, aged from 24-41 (m=28.733, sd=4.079) to test on Google Glass, and then we compared the two groups score. We only compared the designed gestures in the user study instead of the experience of whole applications.

The first stage is comprised of Likert scale questionnaire to reveal the usability of the eleven designed gestures. In the test for the first user group, the mean of usability score indicates that all gestures are applicable. The KW test reveals a significant effect ($\chi^2 = 24.63, p = 0.0061$) on usability for the gestures at $p = .01$ level. In the test for the second user group, the means of usability scores are a little more than that for the first user group except *'Swing Finger Fast'* gesture. The KW test reveals a significant effect ($\chi^2 = 18.64, p = 0.0451$) at $p = .05$ level. It reveals that the designed gestures are not restricted by the device exhaustively. The result indicates that Google Glass almost bring more intuitive and comfortable user experience to the users. The decreased score of *'Swing Finger Fast'* gesture can be explained as that the HWD isn't fixed on the wrist or knee so that the hands' motion can use more joints, thus *'Swing Finger Fast'* gesture entails more strenuous effort. Another appearance is that *'Swing Finger Left'* gets more score than *'Swing Finger Right'* in the first test, but gets the same score in the second test. The reason should be similar: in the first test the wearable framework is fixed on the left wrist thus the users feel more comfortable to swing the hand to the left than right. However, in the second test the available joints are not restricted in the wrist therefore the users don't care about using left or right hand. Wilcoxon signed-rank test result reveals that smart glass is not influential in the designed gestures, in other words, the gestures are suitable on ubiquitous context.

In the second stage, in order to assess the user's emotions, we used the Geneva Emotions Wheel (GEW). GEW allows us to address the pleasantness and control dimensions of emotions and was handed to the users immediately after they conducted the given tasks. The result of Wilcoxon signed-rank test reveals that smart glass is influential in the user's pleasant emotion for using the designed gestures. The situations of emotion test indicate that people feel to be more in control over the emotion than not while the difference is that Google Glass hasn't brought significant improvement for controlling the proposed touch-less gestures. The proposed interaction method with both devices affects the user's emotions significantly. Positive emotions of the GEW are rated slightly more effective than negative emotions.

Concerning the use cases, we have conducted deep user study comparing these unconventional interaction styles. Since the touchless motion interaction hasn't to depend on range sensor or dual camera, so it is suitable for modern smart glasses devices. The touch-less on smart glasses extends the operation space resolution not only because it extends the distance between hands/feet and camera, but also that it completely frees the users hands from the hand-held device, so both hands are available to create more gestures. In addition, the visual feedback from the glasses projector is directed at the right eye, that is intuitive WYSIWYG interaction.

## 4 CONCLUSION

Our solution provides a low-cost solution for the professional application on vision based wearable device. A series of designed gestures and their evaluation related to touch-less interaction technology on vision based wearable device is proposed. The comparing result of the user studies reveal that the smart glass is a preferably platform for our touch-less interaction approach. In future, the proposed interaction approach will have possibility of benefiting various of scientific and civil field, such as 3D molecular interactive visualization [10], 3D seabed visualization [9], 3D virtual reality geographic information system (VRGIS) [7], virtual community [2].

### ACKNOWLEDGEMENT
Authors would like to thank Dr. Alaa Halawani and Dr. Shafiq Ur Rehman.